# Tattoo-Paper Transfer as a Versatile Platform for All-Printed Organic Edible Electronics

*Giorgio E. Bonacchini[1,2], Caterina Bossio[1], Francesco Greco[3,4], Virgilio Mattoli[3], Yun-Hi Kim[5], Guglielmo Lanzani[1,2]\*, Mario Caironi[1]\**

[1] Center for Nano Science and Technology @PoliMi, Istituto Italiano di Tecnologia, Via Giovanni Pascoli, 70/3, 20133 Milano, Italy

[2] Department of Physics, Politecnico di Milano, Piazza Leonardo da Vinci, 32, 20133 Milano, Italy

[3] Center for Micro-BioRobotics @SSSA, Istituto Italiano di Tecnologia, Viale Rinaldo Piaggio 34, 56025 Pontedera, Italy

[4] Department of Life Science and Medical Bioscience, Graduate School of Advanced Science and Engineering, Waseda University, 2-2 Wakamatsu-cho, Shinjuku-ku, Tokyo, Japan

[5] Department of Chemistry and Research Institute of Green Energy Convergence Technology (RIGET), Gyeongsang National University, Jinju, South Korea

**Abstract**

The use of natural or bioinspired materials to develop edible electronic devices is a potentially disruptive technology that can boost point-of-care testing. The technology exploits devices which can be safely ingested, along with pills or even food, and operated from within the gastrointestinal tract. Ingestible electronics could potentially target a significant number of biomedical applications, both as therapeutic and diagnostic tool, and this technology may also impact the food industry, by providing ingestible or food-compatible electronic tags that can "smart" track goods and monitor their quality along the distribution chain. We hereby propose temporary tattoo-paper as a simple and versatile platform for the integration of electronics onto food and pharmaceutical capsules. In particular, we demonstrate the fabrication of all-printed Organic Field-Effect Transistors (OFETs) on untreated commercial tattoo-paper, and their subsequent transfer and operation on edible substrates with a complex non-planar geometry.

**Main text**

Point-of-care testing is defined as the administration of real-time medical diagnostic assays directly near the place of patient care. The driving notion behind point-of-care devices is to bring the test conveniently and immediately to the patient, increasing the likelihood that the physician and the care team will receive the results quicker, thus allowing for rapid and more efficient clinical management decisions. Such scenario has been enabled by the development of portable, simple to use and easy to read medical equipment. With the development of edible devices that can be safely ingested by the patient, perform a test or release a drug and then transmit a feedback, the pervasiveness of point-of-care testing can be further enhanced. The cost-effective and mass-scale production of this technology could favor its adoption also in underdeveloped countries, where many health issues are still endemic within the population, thanks to costs reduction. Furthermore, the food industry may find important applications of this technology too, in tracking, identification or monitoring of food specimens during preparation, delivery and stocking.[1–3] Until now, a number of ingestible technologies have been proposed to semi-invasively perform relevant biomedical therapeutic or diagnostic tasks, including edible power sources and energy harvesters,[4–7] adherence monitoring devices,[8] controlled drug-release systems,[9–11] sensors,[12–14] and even food quality monitoring devices.[15,16] However, the electronic circuitry employed in all these works has so far been based exclusively on standard Si-based electronics, or other inorganic technologies, which are often sizeable and bulky, and necessarily require cumbersome and costly fabrication processes, thus hampering the diffusion of edible electronics. In particular, these are energy intensive fabrication techniques, hardly scalable, and non-ideal for the low-cost high-throughput realization of disposable, one-time-use edible electronics (particularly in food monitoring applications). On the other hand, organic electronic materials hold great potential, thanks to the possibility of realizing electronics at low-cost, with accessible and easily up-scalable techniques.[17–20] Moreover, the ability of "soft" organic materials to provide new interaction paths between electronic devices and living tissues, as well as the possibility to chemically synthesize a plethora of different molecules and polymers, may result in a novel class of edible devices with improved mechanical properties and biocompatibility.[21,22] Nonetheless, despite their strong potential, very few examples of organic electronic devices have been specifically proposed for this kind of applications. Contributions to this field include a set of edible passive wireless antennas able to conform to different foods and monitor their degradation in time,[15] as well as current sources fabricated with edible materials whose performances are compatible with those of conventional power supplies currently adopted for Si-based ingestible devices.[5,23,7] The only example of OFET fabricated on edible substrates was demonstrated by Irimia-Vladu *et al.*, obtained by physical vapor deposition of natural and nature-inspired materials directly onto surface-engineered hard-gelatine capsules and caramelized glucose.[1] Conversely, the use of commercial tattoo-paper for the integration of biocompatible active electronic devices on edible substrates opens new and exciting perspectives in the long-term vision of widely accessible personal care applications: it is compatible with cost-effective and mass-manufacturing printing techniques,[24,25] and it provides a general and flexible platform for easy integration of the devices on many different types of objects, without any type of surface treatment prior to device fabrication or transfer.

Temporary tattoo-paper (or decal transfer paper) consists of a submicrometric film of ethylcellulose (EC), a cellulose derivative, attached to a paper sheet by means of a sacrificial water-soluble starch/dextrin layer (Figure 1a). EC is an ingestible material that can be commonly found in drug formulations as a coating agent, tablet binder or filler, while it is used as emulsifier in the food industry.[26] By exploiting tattoo-paper, it is possible to easily and reliably transfer the submicrometric ethylcellulose layer onto many different items: the tattoo-paper is soaked with water that dissolves the starch sacrificial layer, and it is then pressed onto the target object; finally, the paper sheet is peeled off to release the conformable, hundreds of nanometers thick EC layer. In this work, different sets of p-type and n-type OFETs were fabricated onto temporary tattoo-paper, where EC acted both as transferrable substrate and as gate dielectric layer of the OFETs.

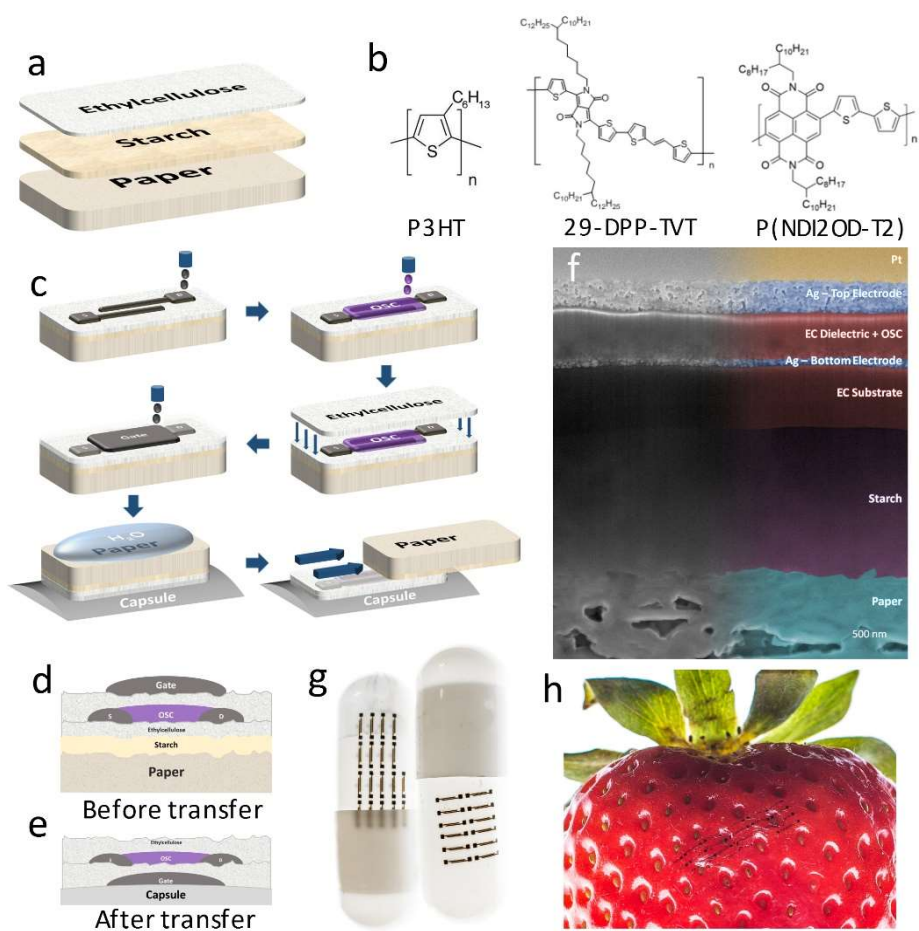

**Figure 1.** OFET structure and fabrication on tattoo-paper. (a) Scheme of the tattoo-paper stack: EC is attached to standard paper by a sacrificial water-soluble starch/dextrin layer (thicknesses not to scale). (b) Molecular structure of the semiconductors. (c) Process-flow of the OFET

fabrication on tattoo-paper. (d, e) Schematic representation of the final structure of the device before (d) and after (e) transfer. (f) Scanning electron microscope (SEM) image of a cross section, milled by focused ion beam (FIB), of the device on tattoo paper: it is possible to distinguish the layers composing the substrate, EC, starch/dextrin and paper, as well as the sintered AgNPs bottom and top contacts, and the dielectric EC layer. (g, h) Photograph of a set of silver electrodes transferred on to a pharmaceutical capsule (g) and on a strawberry (h).

OFET electrodes were realized by inkjet-printing and subsequent sintering of a commercially available silver nanoparticle (AgNP) ink. While bulk silver is an established bioinert food garnish, with a Recommended Dietary Allowance (RDA) of 350 $\mu g_{Ag}$/day for a 70 kg person, the assumptions on the AgNP biocompatibility with the human body are based on previous works.[2,5,27] and on the fact that sintered nanoparticles create a continuous layer of material that hinders phagocytosis, which is a dominant mechanism for in vitro cytotoxicity. As additional arguments, the amount of material used for a single transistor is in the range of 4 $\mu g_{Ag}$, well below the suggested limits for human daily consumption even in the case of simple, small-scale of integration circuits.[28]

Four different polymer semiconductors were used in our tests to demonstrate the versatility of the transfer procedure and the compatibility with both good hole and electron transporting materials, as required by robust complementary logic circuits.[19] A well-known and biocompatible material, poly(3-hexylthiophene) (P3HT),[29,30] was employed both as pristine polymer and in a semiconductor-insulator polymer blend configuration with polystyrene (PS), in order to enhance and stabilize the performances of the OFETs.[31] We also employed some recently developed, high-performance donor-acceptor copolymers, such as poly[2,5-bis(2-decylnonadecyl) pyrrolo[3,4-c]pyrrole-1,4-(2H,5H)-dione-(E)-1,2-di(2,2′-bithiophen-5yl)ethene] (29-DPP-TVT) for p-type devices and poly{[N,N′-bis(2-octyldodecyl)-naphthalene-1,4,5,8-bis(dicarboximide)-2,6-diyl]-alt-5,5′-(2,2′-bithiophene)} (P(NDI2OD-T2)) for n-type devices. These state-of-the-art materials have reported field-effect hole and electron mobilities exceeding 1 $cm^2V^{-1}s^{-1}$ respectively, depending on the deposition technique.[32,33] While testing the tattoo-transfer methodology with high-performance donor-acceptor co-polymers is a relevant choice to assess our approach, no account on the biocompatibility of 29-DPP-TVT and P(NDI2OD-T2) is currently available, to the best of our knowledge. In spite of the extremely low quantity of active material exploited for the fabrication of a single OFET (in the range of picograms, see Table S1), this poses an obvious question on the overall biocompatibility of the final devices, which we address with preliminary experiments in a dedicated section.

The device fabrication steps are sketched in **Figure 1c**. The silver source and drain electrodes are inkjet-printed and subsequently sintered directly on untreated tattoo-paper, forming a channel with a width (*W*) of ~ 1000 μm and a length (*L*) of ~ 50 μm. Likewise, the semiconductor is also inkjet-printed, and annealed to remove residual solvent from the film. A separate piece of tattoo-paper is then used to laminate a thin EC layer on the device, by exploiting the already mentioned transfer technique. This second EC layer effectively acts as gate dielectric, providing an average gate capacitance of approximately 5.5 nFcm$^{-2}$ at 100 Hz (Figure S2), a value compatible with conventional low-k insulating polymers such as PMMA.[34] At the end of the fabrication process, as shown in **Figure 1d**, the resulting OFET has a staggered top-gate bottom-contact configuration, and lies on tattoo-paper ready for transfer. The latter is achieved by first soaking the entire sample into water to dissolve the starch sacrificial layer, and then by placing it onto the edible substrate and removing the paper. OFETs are transferred on two different types of substrates: a glass microscope slide, serving as a rigid and planar reference, and a pharmaceutical, hard-gelatine capsule bought in a local pharmacy. In order to remove the water in excess after the final transfer, without compromising the integrity of the capsules, all samples are left to dry overnight in vacuum at pressure not inferior to 10 mbar, as such level is currently compatible with commercial vacuum drying (vacuum oven) technology, largely employed in the food and pharmaceutical industry.[35] Overall, the fabrication procedure encompasses only the use of scalable printing techniques and the tattoo transfer process.

Electrical characterization of the samples was performed before and after transfer on the target substrate, to verify, through the comparison of results, eventual effects on the electrical performances upon lamination. Each set of OFETs, i.e. samples intended for transfer on glass and on capsule separately, consisted in a number of devices ranging from 15 to 25 per semiconductor type, to allow for an appropriate evaluation of the transfer procedure robustness (see Table S2 and Table S3, reporting respectively the transfer yields and the statistics on device parameters, including average $V_t$, average and highest mobilities, average *ON/OFF* and maximum currents in saturation regime).

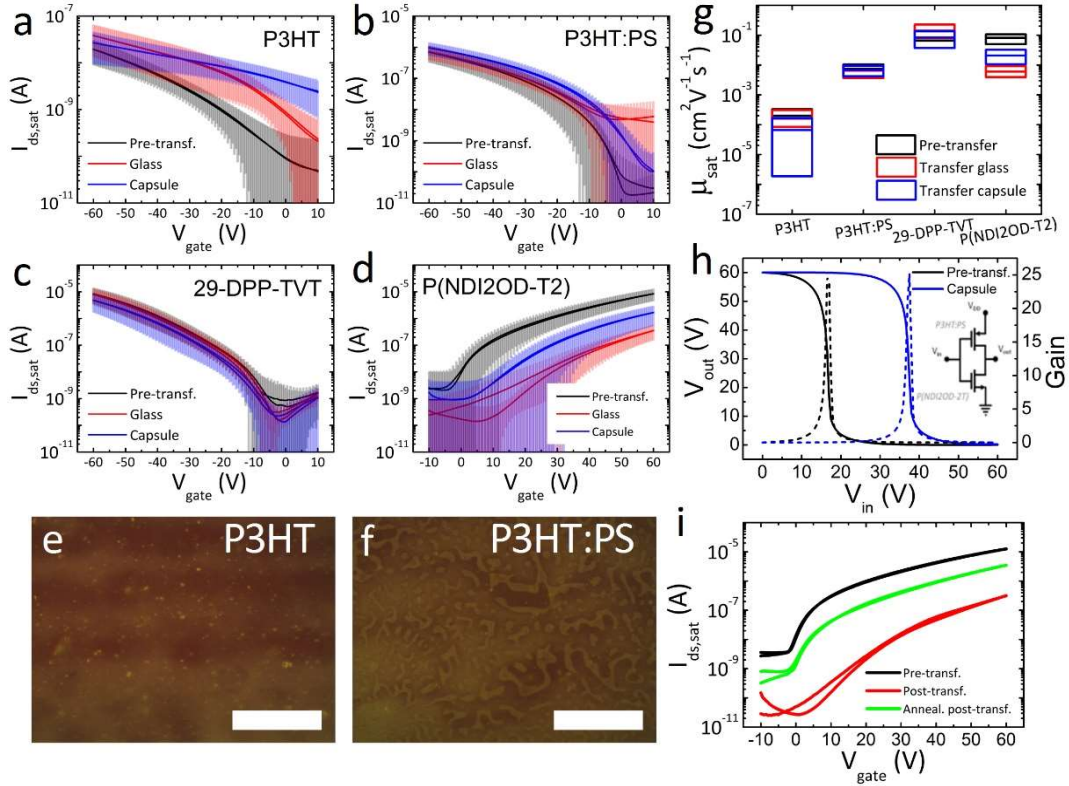

**Figure 2.** OFET and logic inverter electrical characterization. (a-d) Average transfer characteristic curves in saturation regime ($V_{ds}$ = -60 V) of OFETs before (black) and after the transfer on glass (red) and on a pharmaceutical capsule (blue), for the four different active layers: neat P3HT (a), P3HT:PS blend (1:1) (b), 29-DPP-TVT (c) and P(NDI2OD-2T) (d); shaded areas represent Standard Deviation. OFETs geometrical parameters: $W \approx 1000$ μm and $L \approx 50$ μm. (e, f) Dark-field microscope images of printed P3HT (e) and P3HT:PS (f) films on EC, with scale-bar corresponding to 50 μm. (g) Average saturation mobilities for the four different semiconductors, before and after transfer, where the boxes indicate Standard Deviation. (h) VTC and gain curve of complementary inverter fabricated with P3HT:PS and P(NDI2OD-2T), acquired before (black) and after transfer on a pharmaceutical capsule (blue); (i) Saturation transfer curves of sample n-type device before (black) and after transfer on glass (red), with green curve demonstrating partial recovery of initial device performance after annealing in inert atmosphere.

In **Figure 2**, the average OFET saturation currents and their standard deviation, obtained before and after transfer, as a function of the gate-source voltage ($V_{gs}$), are presented for the four different active layers. Devices prepared with pristine P3HT after the transfer are subject to an evident increase of the off current and shift of the threshold voltage to positive values, with average hole mobilities stable around $10^{-4}$ cm$^2$V$^{-1}$s$^{-1}$. As previously described in literature, this behaviour suggests that the transfer procedure, which exposes the devices to light, air and water, produces a doping of the active layer, which is not effectively reversed by the vacuum treatment.[36] Several works have investigated the improved performances of polythiophene-based semiconductors when blended with insulating polymers, highlighting their enhanced mechanical flexibility, the superior transport properties and environmental stability.[31,36] In order to improve the stability of our devices' performances, P3HT was blended with PS in a 1:1 weight ratio and deposited following the same procedure. Microscope images of the active layer inkjet-printed on EC (Figures 2e, 2f) show the difference in morphology of the pristine semiconductor compared to the P3HT:PS blend. The latter, differently from the homogeneity of the pristine P3HT film, presents the features typical of a phase-separated sea-island morphology, the darker area likely constituting a percolative semiconducting path through PS islands.[37] The positive impact of blending on the device performances is evinced by an average increase of hole mobility of almost two orders of magnitude, consistent with what obtained by Lu G. *et al.*, who witnessed an enhanced crystalline order of P3HT when embedded in an inert matrix. The characteristic transfer curves appear more robust to the transfer procedure, especially in full-accumulation regime, with hole mobilities of approximately $7 \times 10^{-3}$ cm$^2$V$^{-1}$s$^{-1}$ before as well as after transfer. The curves show only a modest variation in threshold voltage, particularly in the case of devices transferred on the pharmaceutical capsule, while those transferred on glass are mainly affected by an increase in the off current. Both of these effects are again to be ascribed to the undesired and uncontrolled doping of P3HT during the transfer procedure, although in this case the pre-transfer characteristics in the on state of the OFET are largely preserved as a consequence of semiconductor encapsulation in the insulating matrix. Further mechanisms that contribute to the performance integrity after transfer are possibly the vertical composition gradient of the blend which limits the off current levels, as well as the improved molecular packing of the semiconductor.[36]

To verify that modifications in P3HT characteristic curves are an effect of ambient instability rather than of the transfer process, we fabricated a third set of devices using a more stable semiconductor. For the purpose we selected a diketopyrrolopyrrole based co-polymer, 29-DPP-TVT, which thanks to the lower HOMO level, lying at -5.26 eV as reported by Yu H. *et al.*,[32] is inherently less prone to ambient oxidation. The impact of the transfer process on the performance of the devices is indeed rather limited for the 29-DPP-TVT case compared to the pristine P3HT one. The average mobility values are stable in the range of 0.1 cm$^2$V$^{-1}$s$^{-1}$, with a modest decrease in the case of devices transferred on pharmaceutical capsule, and the threshold voltages do not vary significantly before and after transfer.

Conversely, the transfer process has an appreciable effect on n-type devices fabricated with P(NDI2OD-2T). Figure 2d demonstrates a marked positive shift of the threshold voltage as a consequence of the transfer, as well as a reduction of the average saturation mobility, from $7 \times 10^{-2}$ cm$^2$V$^{-1}$s$^{-1}$ to approximately $7 \times 10^{-3}$ cm$^2$V$^{-1}$s$^{-1}$, and a less sharp turn on region. All of these effects are well in agreement with the degradation mechanisms proposed by Di Pietro *et al.*, which are induced by oxygen and water impurities interacting, both reversibly and irreversibly, with the NDI section of the copolymer backbone.[38] Indeed, partial recovery of the OFETs performances was achieved on those devices transferred on glass and annealed in inert atmosphere for approximately two hours (Figure 2i). We thus identify the cause of the substandard performances in the presence of environmental impurities that are not effectively removed by the vacuum treatment alone. Similarly to the p-type case, the selection of an inherently more environmentally stable electron transporting material should produce unmodified characteristic curves upon transfer.

Our experiments overall demonstrate the possibility to transfer both p- and n-type transistors thanks to a paper-based temporary tattoo, where the final performance of the transferred device is dominated by the intrinsic ambient stability of the adopted semiconductor. This result paves the way for the realization of robust complementary circuits: to this extent, we employed the same process to fabricate a complementary logic inverter. Figures 2g and 2h report the inverter voltage transfer curves (VTC) obtained using P3HT:PS and P(NDI2OD-2T) for the p-type and n-type devices, respectively, on both glass and pharmaceutical capsule (noise margins and gain values reported in Table S4). As previously described, the n-type device is subject to a decrease in average maximum current of about one order of magnitude. For this reason, the inverter was deliberately designed to appear unbalanced before the transfer, the n-type OFET having geometrical parameters equal to the p-type transistor (both $W \approx 1000$ μm and $L \approx 50$ μm), so to outperform it before the transfer and thus obtain a more balanced inverter after lamination. In all cases, from the VTC, we extracted average gain values ranging from 14 to 25.

The good electronic performance of the transferred circuits is meaningful if the adopted semiconductors are compatible with ingestion. Such demonstration is a complex task that would eventually require clinical tests. To start preparing the ground for future trials, we evaluated the possible cytotoxicity of 29-DPP-TVT and P(NDI2OD-2T) by means of the 3-(4,5-dimethylthiazol-2-yl)-2,5-diphenyltetrazoliumbromide (MTT) assay. We performed MTT assays on HUTU-80 and CACO-2 cells after 1, 2, 3 and 4 days *in vitro*. The two cell lines were specifically chosen for their upper and lower intestine derivation, being respectively from human duodenum and epithelial colorectal adenocarcinoma, and thus providing first indications on the impact of the organic semiconductors on the GI tissue (P(NDI2OD-2T) also tested with HEK-293 cell line, human embryonic kidney cells, see SI). The MTT used for the proliferation assay is reduced to formazan solely in living cells, the reduction depending on the cellular metabolic activity, hence stronger formazan optical absorption indicates progressively larger cell population. **Figure 3a** and 3b show the results of the MTT assay carried with both cell lines on the two different semiconducting films, demonstrating that cell proliferation is substantially unaffected by the presence of the semiconducting polymers, thus excluding biocompatibility issues of the materials at this stage.[39,40] In order to ensure that cellular morphology is not altered when cells adhere to the semiconducting polymers, both HUTU-80 and CACO-2 were stained with phalloidin conjugated to a fluorophore, which selectively binds to f-actin proteins in living cells, and analysed by fluorescence microscopy. Figure 3c shows that the two cell lines present the same cellular morphology when grown on either of the semiconducting polymers and the glass control substrate. It must be noted that this biocompatibility analyses do not however cover other complex physiologic reactions that could be triggered by the device at the tissue and organ levels, which will require dedicated studies.

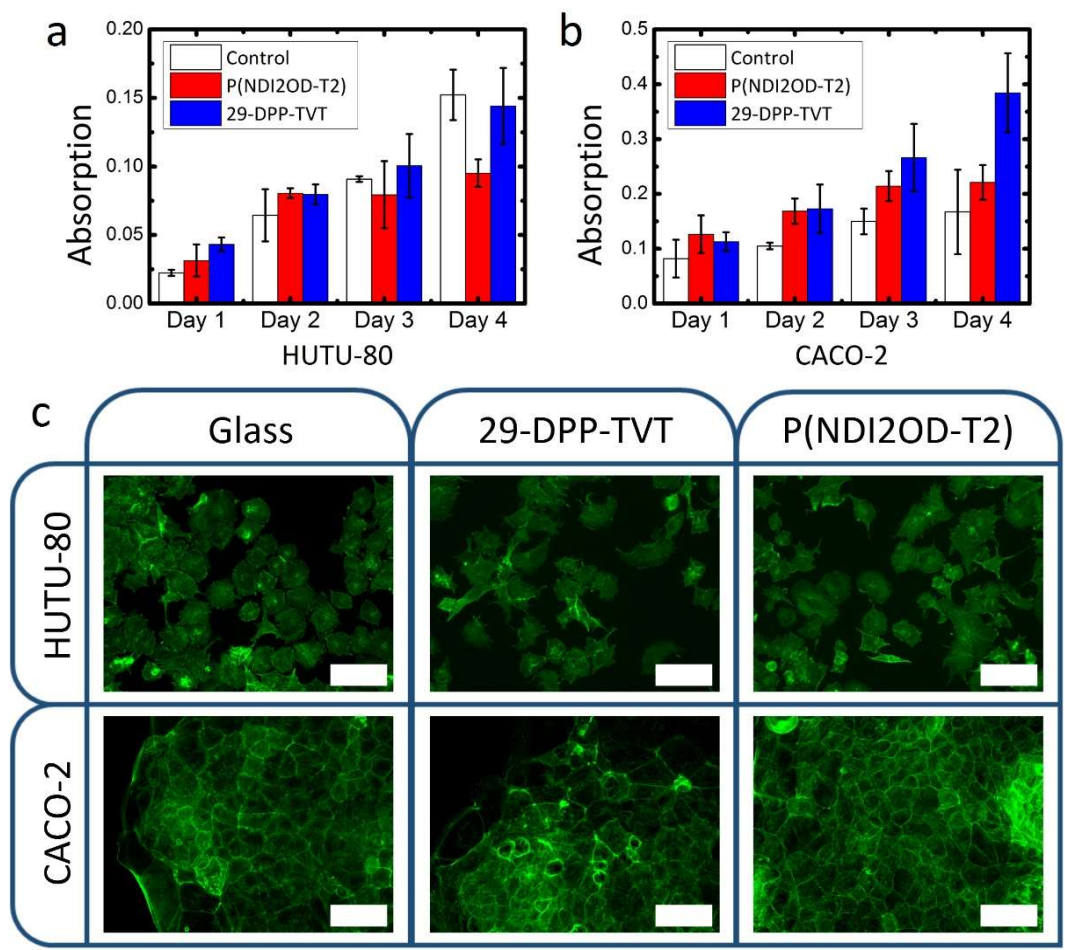

**Figure 3.** Cell proliferation assays on high-performance organic semiconductors. Optical absorption of formazan collected at 1, 2, 3 and 4 days *in vitro* (a, b) and inverted fluorescence

microscope images (c) reported for HUTU-80 and CACO-2 cell lines grown on 29-DPP-TVT, P(NDI2OD-2T) and on glass as control. Scale-bars in 3c correspond to 100 μm. Error bars in (a, b) correspond to Standard Error estimated on three samples.

In conclusion, organic transistors composed mainly of ingestible materials, both p-type and n-type, can be easily transferred on edible substrates by means of untreated commercial tattoo-paper. This system constitutes a simple and versatile platform for the integration of fully printed organic circuitry on food and pharmaceutical drugs, and it relies on a transfer procedure that does not alter the integrity of the ingestible substrate. The robustness of the transfer is evaluated on a statistically relevant number of OFETs, employing one biocompatible polymer and two high-mobility co-polymer semiconductors. A preliminary degree of biocompatibility of the former two is here assessed for the first time by cell proliferation assays, performed with two different cell lines derived from human GI tissue.

Future works will focus on improving the overall performance stability of the device and its active layer, as well as on an aggressive scaling of the driving voltages in order to support OFETs' operation at power levels compatible with state-of-the-art edible batteries. Interestingly, much higher specific capacitance values have been achieved by Petriz *et al.* with a different, specifically engineered, cellulose-based material, hence suggesting that future developments of the tattoo-paper approach could in principle allow for low voltage operation of the OFETs.[41] The realization of such stable, low-voltage, low-cost, easily transferrable, biocompatible organic electronic devices will pave the way to the development of a new set of point-of-care testing systems, constituted by edible sensors able to effectively operate from within the GI tract. Ingestible electronic systems that can be degraded within the digestive tract are attractive with respect to passivated inorganic electronics, as they completely remove the possibility of accumulation in the human body, which may become an issue in case of severe pathologies. Inherent edibility may also enable the direct integration of electronic sensors into food in the form of embedded smart probes, following a perishable good along its entire lifespan and capable of alerting the consumer when alteration of the properties is taking place. Furthermore, being edibility an extreme case of ecocompatibility, the future development of such technology platform will at the same time favor the reduction of the environmental impact of large area and flexible electronics technologies at large. Overall, with our work we take the first step along a promising path and demonstrate that printed polymer electronics is a potential enabler of such an appealing future technology.

**Experimental Section**

*Materials*: Commercial tattoo-paper (Tattoo 2.1) was acquired from The Magic Touch Ltd. and used without any cleaning or surface treatment. Semiconducting polymers regioregular P3HT (average $M_n$ 54,000-75,000, electronic grade, 99.995% trace metals basis) and P(NDI2OD-2T) (Activink™ N2200) were purchased from respectively Sigma Aldrich and Flexterra Inc., and used as received. 29-DPP-TVT was synthesized according to a previously reported protocol.[24] Polystyrene (analytical standard, average $M_w$ = 290,000, average $M_n$ = 130,000) was acquired from Sigma Aldrich. Silver-nanoparticle ink (TEC-IJ-060) was purchased from InkTec. 1,2-Dichlorobenzene (DCB, anhydrous, 99%) and Mesitylene (Mesy, 98%) were purchased from Sigma Aldrich and used as received.

*Sample Preparation and Transfer*: P3HT, P3HT:PS (1:1 mass ratio) and 29-DPP-TVT were dissolved in DCB (P3HT and P3HT:PS at 8 mg/ml, 29-DPP-TVT at 5 mg/ml) by stirring the solution at ~100 °C for 12 h. These inks were regularly stirred at ~100 °C for 20 min before printing. P(NDI2OD-2T) was dissolved in Mesy (8 mg/ml) by stirring the solution overnight at room temperature.

AgNP ink was inkjet-printed (Fujifilm Dimatix, DMP2831) directly on commercial-tattoo paper without any type of surface treatment or cleaning process, apart from nitrogen flushing to remove dust from the sample area. AgNP ink was sintered immediately after deposition at ~125 °C for 10 minutes in air, to achieve a channel width $W$ of 1000 μm and length $L$ of 50 μm. The semiconductors were inkjet-printed directly on the contact and channel area with the DMP2831, and annealed at ~125 °C for 10 min in air. An additional piece of tattoo-paper was soaked in MilliQ water and deposited on top of the contacts and the active layer area to release the EC layer, thus constituting the gate dielectric of the devices; water in excess was removed by placing the sample at ~125 °C for 10 min in air. Subsequently, the AgNP ink was inkjet-printed on the EC layer acting as gate dielectric and immediately sintered at ~125 °C for 10 min in air. The devices were then annealed in inert atmosphere at ~125 °C (nitrogen filled glovebox, $O_2$ and $H_2O$ levels below 1 ppm) for 2 h (29-DPP-TVT) and 12 h (P3HT, P3HT:PS and P(NDI2OD-2T)). The complementary inverters were fabricated employing the same procedure as the individual devices. All inkjet-printing was performed with a Fujifilm Dimatix DMP2831 printer. Transfer was performed by soaking the tattoo-paper substrate in MilliQ water and subsequent deposition on microscope glass slide and hard gelatin capsule. Water in excess was removed by placing samples in vacuum atmosphere at a minimum nominal pressure of 10 mbar (Edwards RV12 pump) for 12 h.

Samples for cell viability tests were prepared using the same formulations of P3HT, 29-DPP-TVT and P(NDI2OD-2T) used for inkjet-printing. OSCs were spin-coated on clean $SiO_2$ microscope slides rinsed in an ultrasonication bath in deionized water, Acetone and 2-Propanol (10 min each).

*Film and Device Characterization*: Dark-field microscope images of P3HT and P3HT:PS inkjet-printed films were taken in reflection mode with a Zeiss Axio Scope A1.

FIB-milled cross sections and SEM imaging of OFET device on tattoo paper before transfer were obtained with a Dual Beam FIB/SEM Helios Nano-Lab 600i (FEI) equipped with a Field Emission source. Samples surface was prepared by depositing a thin Au film (t ~ 20 nm) by means of a Q150R sputter coater (Quorum Technologies). FIB-assisted deposition of Pt (~25 x 2 x 1 $\mu m^3$) was carried out on the area of interest before FIB-milling and polishing of cross-section (Ga+ source, 30 kV acceleration voltage). SEM images of cross sections were obtained under a sample tilt angle of 52° and with accelerating voltage of 18 kV. In order to highlight the features SEM images were post-processed (colorized) with Adobe Photoshop CS4.

Measurements of the OFETs characteristic curves and of the inverters VTC were performed in inert atmosphere (nitrogen filled glovebox, $O_2$ and $H_2O$ levels below 1 ppm) by means of an Agilent B1500A Semiconductor Parameter Analyzer. Device parameters were extracted considering the characteristic equation for field-effect transistors operating in saturation: $I_{ds}=(WC'_p/2L)\mu_{sat}(V_{gs}-V_t)^2$, where $I_{ds}$ is the source-drain current, $C'_p$ is the capacitance per unit area (5.3 nF/cm$^2$ at 100 Hz) of the EC dielectric layer (see Supplementary Information, Figure S2), $\mu_{sat}$ the carrier mobility in saturation regime, $V_{gs}$ and $V_t$ the gate-source voltage and the threshold voltage respectively. Transfer characteristic curves of p-type (n-type) devices in linear and saturation regimes were acquired at $V_{ds}$ = -5 V ($V_{ds}$ = 5 V) and $V_{ds}$ = -60 V ($V_{ds}$ = 60 V) respectively. Mobility values and threshold voltages were obtained by a linear fitting of the square-root of $I_{ds}$, over a voltage range of approximately 20 V, between 40 V and 60 V. Due to the non-ideal features of the transfer characteristics (see Supplementary Information), such obtained mobilities and threshold voltages should be regarded as OFET device parameters rather than values related to intrinsic properties of the active layers. After transfer, a certain number of devices displayed a non-negligible gate-drain current in the linear regime; only OFETs with leakage current significantly lower than $I_{ds}$ (at least two orders of magnitude) were considered for the statistical evaluation.

The recovery of the performances of n-type devices is obtained by annealing the transferred samples in inert atmosphere at ~125 °C for 2 hours (Figure 2i).

EC capacitance per unit area was extracted in a two probes configuration with an Agilent 4294A Impedance Analyzer. Samples were fabricated by transferring a EC layer on a SiO$_2$ substrate with an evaporated Cr-Au (Cr: 2 nm, Au: 50 nm) bottom electrode and an inkjet-printed AgNP top electrode. Capacitors were annealed in inert atmosphere at ~125 °C for 12 h before the measurements.

*Cell Viability*: In order to evaluate 29-DPP-TVT and P(NDI2OD-2T) biocompatibility, the MTT [3-(4,5-dimethylthiazol-2-yl)-2,5-diphenyltetrazolium bromide] (Sigma Aldrich) assay was performed with two different cell lines: HUTU-80 (from human duodenum adenocarcinoma) and CACO-2 (from human epithelial colorectal adenocarcinoma) (both acquired from American Type Culture Collection).

Cells were cultured in cell culture flasks containing Dulbecco's modified Eagle's medium (DMEM) with 10% Fetal Bovine Serum (FBS), 100 µg ml$^{-1}$ Streptomycin, 100 U ml$^{-1}$ Penicillin and 100 U ml$^{-1}$ L-Glutamine. Culture flasks were maintained in humidified atmosphere at 37 °C with 5 % $CO_2$.

When at confluence, the cells were enzymatically dispersed using trypsin-EDTA and then plated on the different polymer substrates at a concentration of 10,000 cells cm$^{-2}$.

The proliferation was evaluated after 1, 2, 3 and 4 days *in vitro*. For each time point the medium was removed and replaced with RPMI without phenol red containing 0.5 mg ml$^{-1}$ of MTT reagent. Cells were re-incubated at 37 °C for 3 h. Formazan salt produced by cells through reduction of MTT was then solubilized with 200 ml of ethanol and the absorbance was read at 560 nm and 690 nm (using a microplate reader TECAN Spark10M). The proliferation cell rate was calculated as the difference in absorbed intensity at 560 nm and 690 nm.

*Immunofluorescence Staining*: Cells grown on glass coverslips coated with 29-DPP-TVT and P(NDI2OD-2T) were washed twice with PBS and fixed for 15 min at RT in 4 % paraformaldehyde and 4 % sucrose in 0.12 M sodium phosphate buffer, pH 7.4. Fixed cells were pre-incubated for 20 min in gelatin dilution buffer (GDB: 0.02 M sodium phosphate buffer, pH 7.4, 0.45 M NaCl, 0.2% (w/v) gelatin) containing 0.3% (v/v) Triton X-100, and subsequently incubated with Phalloidin Alexa Fluor 488 conjugated in GDB for 1 h at RT and finally washed with PBS. The images were acquired with an inverted fluorescence microscope (Nikon Eclipse Ti-U/ Lumencor Spectra X).

**Acknowledgements**
The authors thank A. Luzio for the photographs of capsules and strawberries, A. D. Scaccabarozzi for discussions on the polymer semiconductor:insulator blend formulation, F. Fumagalli for the insights on vacuum technology, M. Garbugli, N. Piva, A. Luzio, G. Dell'Erba and M. R. Antognazza for useful discussions. M. C. acknowledges support by the European Research Council (ERC) under the European Union's Horizon 2020 research and innovation program 'HEROIC', grant agreement 638059. F. G. acknowledges financial support from Top Global University Project at Waseda University, Tokyo from MEXT Japan.

Supporting Information

**Tattoo-Paper Transfer as a Versatile Platform for All-Printed Organic Edible Electronics**

*Giorgio E. Bonacchini, Caterina Bossio, Francesco Greco, Virgilio Mattoli, Yun-Hi Kim, Guglielmo Lanzani\*, Mario Caironi\**

| Materials | Dose [grams per device] |
|---|---|
| P3HT | $2.5 \times 10^{-12}$ (pg) |
| P(NDI2OD-T2) | $2.5 \times 10^{-12}$ (pg) |
| 29-DPP-TVT | $1.6 \times 10^{-12}$ (pg) |
| P3HT:PS | $2.5 \times 10^{-12}$ (pg) |
| Sintered AgNPs | $4 \times 10^{-6}$ (μg) |
| Ethylcellulose | $6.4 \times 10^{-6}$ (μg) |

**Table S1.** Estimated amount of semiconducting and conductive material for each OFET in grams per device (units in brackets indicate order of magnitude).

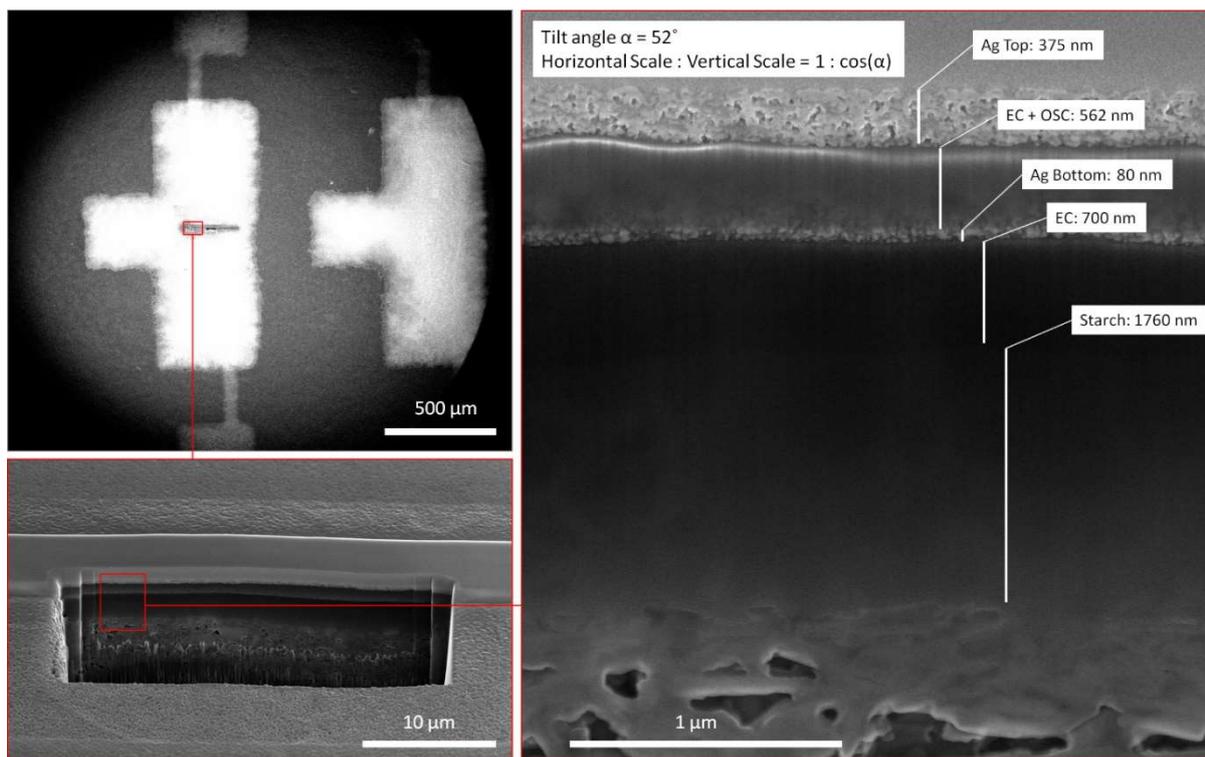

**Figure S1.** SEM images of OFET on tattoo-paper, before transfer, and its cross section produced by FIB milling. *Left, top*: top view: *left, bottom*: tilted view (α = 52˚) of milled cross section area. *Right:* high magnification tilted view of cross section with indications regarding the thickness of all the layers.

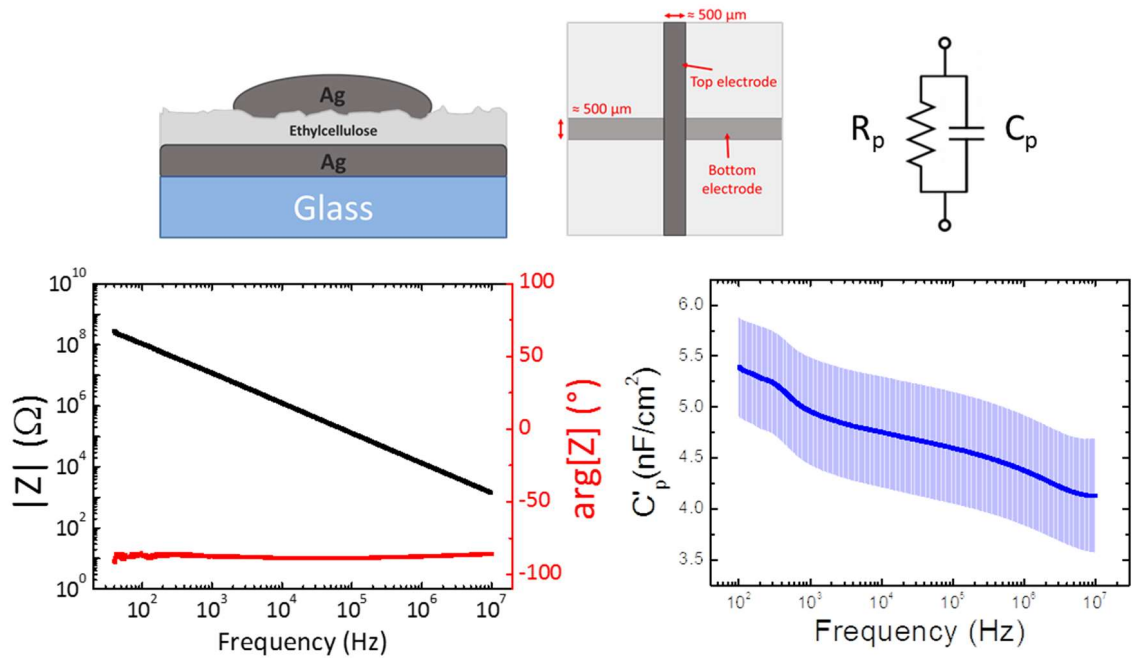

**Figure S2.** Top images report on EC capacitor structure, and on equivalent electrical circuit used to extract effective $C'_p$ from impedance data. Bode plots of characteristic device (bottom-left) and average effective specific capacitance versus frequency (top-right) of EC. Shaded area represents standard deviation of average $C'_p$, calculated from 7 devices.

|  | P3HT | P3HT:PS | 29-DPP-TVT | P(NDI2OD-2T) |
|---|---|---|---|---|
| Pre-transfer | 42/50 | 47/50 | 47/50 | 39/41 |
| Transfer on glass | 17/22 | 24/24 | 24/24 | 12/15 |
| Transfer on capsule | 18/20 | 22/23 | 16/23 | 17/24 |

**Table S2.** Number of devices working for each type of active material, before and after transfer on glass and on capsule.

|  | Pre-transfer | | | | |
|---|---|---|---|---|---|
|  | $\mu_{avg}$ [cm$^2$V$^{-1}$s$^{-1}$] | $\mu_{max}$ [cm$^2$V$^{-1}$s$^{-1}$] | $V_t$ [V] | ON/OFF | $I_{max}$ [A] |
| P3HT | (1.97±1.31)x10$^{-4}$ | 5.90x10$^{-4}$ | -11.43 | 10$^2$ | 1.94x10$^{-8}$ |
| P3HT:PS | (7.09±3.29)x10$^{-3}$ | 1.51x10$^{-2}$ | -18.10 | >10$^4$ | 6.95x10$^{-7}$ |
| 29-DPP-TVT | (1.46±0.81)x10$^{-1}$ | 3.24x10$^{-1}$ | -28.61 | 10$^4$ | 8.31x10$^{-6}$ |
| P(NDI2OD-2T) | (7.83±2.87)x10$^{-2}$ | 1.38x10$^{-1}$ | 16.48 | 10$^4$ | 8.38x10$^{-6}$ |
|  | Transfer on glass | | | | |
|  | $\mu_{avg}$ [cm$^2$V$^{-1}$s$^{-1}$] | $\mu_{max}$ [cm$^2$V$^{-1}$s$^{-1}$] | $V_t$ [V] | ON/OFF | $I_{max}$ [A] |
| P3HT | (1.97±1.12)x10$^{-4}$ | 4.18x10$^{-4}$ | -1.16 | 10$^2$ | 3.82x10$^{-8}$ |
| P3HT:PS | (6.42±2.73)x10$^{-3}$ | 1.09x10$^{-2}$ | -14.64 | 10$^2$ | 7.13x10$^{-7}$ |
| 29-DPP-TVT | (1.53±0.73)x10$^{-1}$ | 2.73x10$^{-1}$ | -30.04 | 10$^4$ | 7.97x10$^{-6}$ |
| P(NDI2OD-2T) | (6.54±2.59)x10$^{-3}$ | 1.17x10$^{-2}$ | 29.75 | 10$^3$ | 3.53x10$^{-7}$ |
|  | Transfer on capsule | | | | |
|  | $\mu_{avg}$ [cm$^2$V$^{-1}$s$^{-1}$] | $\mu_{max}$ [cm$^2$V$^{-1}$s$^{-1}$] | $V_t$ [V] | ON/OFF | $I_{max}$ [A] |
| P3HT | (8.4±8.21)x10$^{-5}$ | 3.03x10$^{-4}$ | 46.90 | 10 | 2.68x10$^{-8}$ |
| P3HT:PS | (6.83±2.61)x10$^{-3}$ | 1.07x10$^{-2}$ | -8.95 | 10$^4$ | 9.54x10$^{-7}$ |
| 29-DPP-TVT | (8.70±0.50)x10$^{-2}$ | 1.68x10$^{-1}$ | -29.48 | 10$^4$ | 4.77x10$^{-6}$ |
| P(NDI2OD-2T) | (2.16±1.10)x10$^{-2}$ | 5.34x10$^{-2}$ | 23.83 | 10$^3$ | 1.67x10$^{-6}$ |

**Table S3.** Statistics on device parameters at before and after the transfer process.

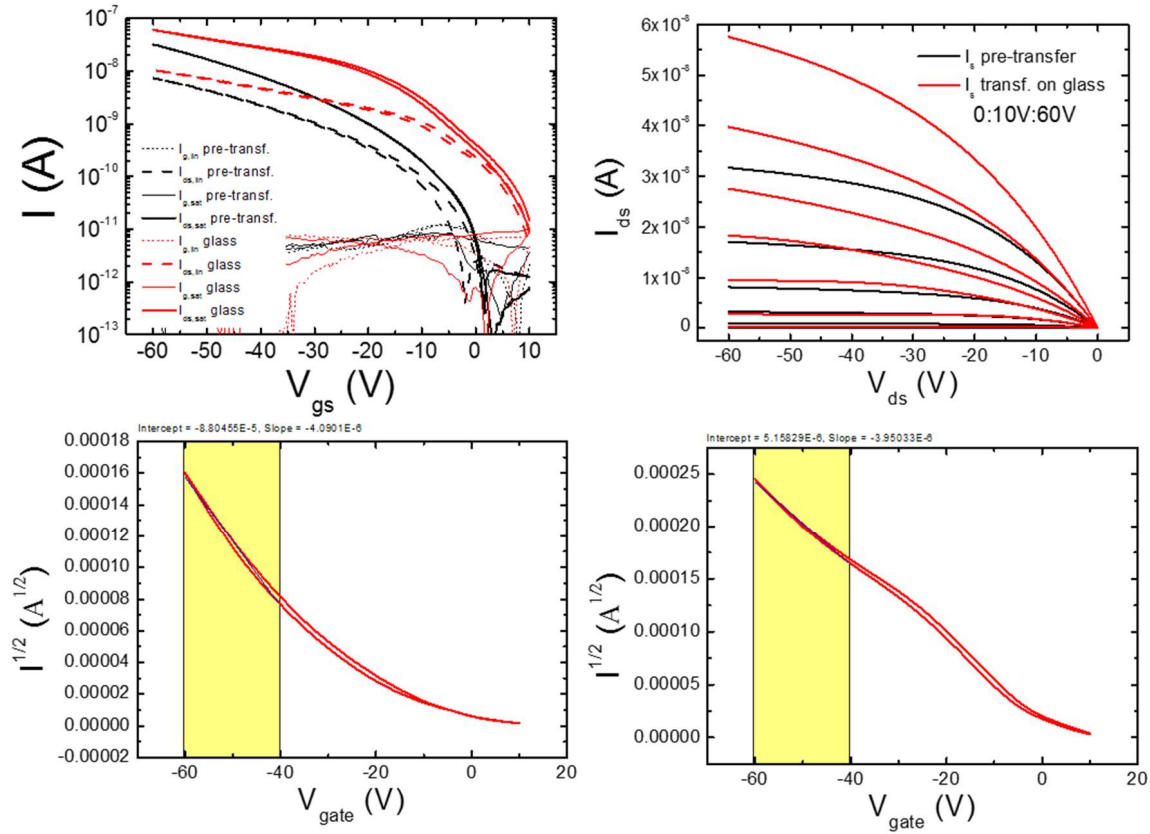

**Figure S3.** Sample OFET fabricated with P3HT and transferred on glass. Transfer curves (top-left) obtained with $V_{ds,lin}$ = -5 V and $V_{ds,sat}$ = -60 V. Output curves (top-right) curves obtained with $V_{gs}$ steps of -10 V between 0 V and -60 V. Bottom images represent $\sqrt{I_{ds}}$ curves for the extraction of mobility and threshold voltage, before (left) and after (right) transfer.

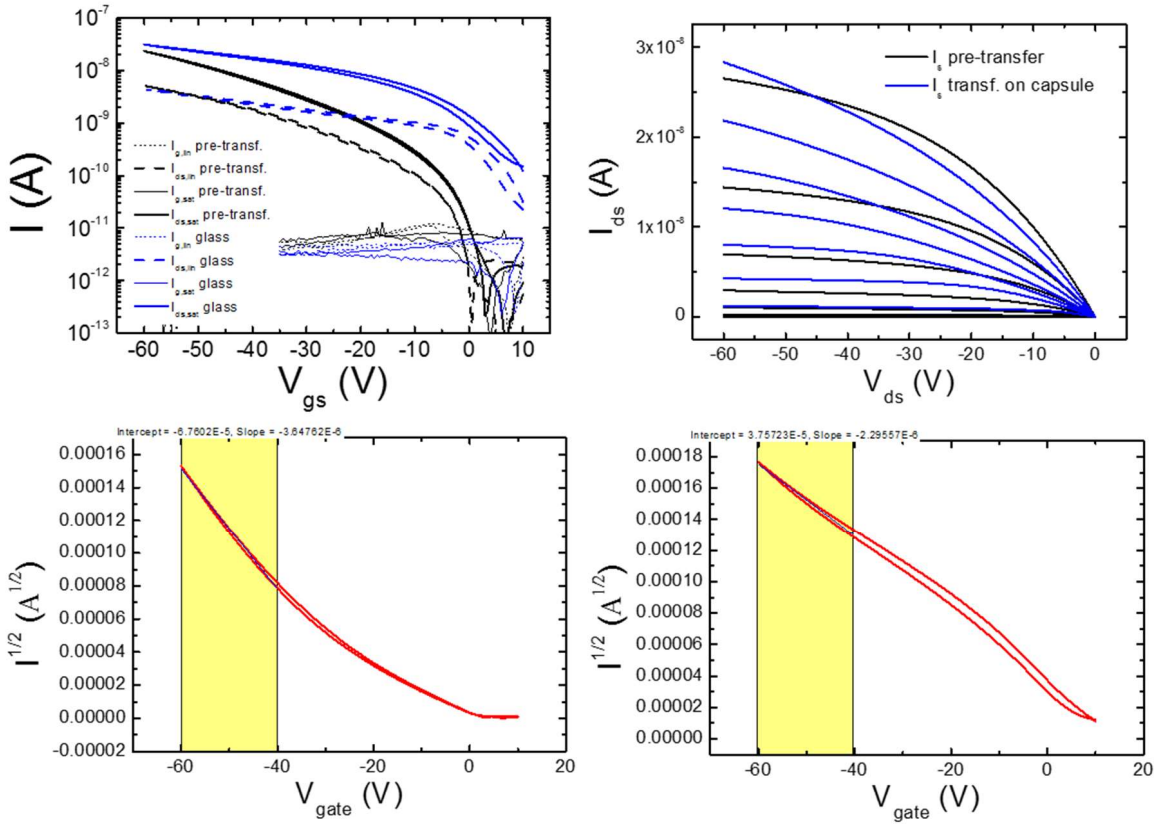

**Figure S4.** Sample OFET fabricated with P3HT and transferred on capsule. Transfer curves (top-left) obtained with $V_{ds,lin}$ = -5 V and $V_{ds,sat}$ = -60 V. Output curves (top-right) curves obtained with $V_{gs}$ steps of -10 V between 0 V and -60 V. Bottom images represent $\sqrt{I_{ds}}$ curves for the extraction of mobility and threshold voltage, before (left) and after (right) transfer.

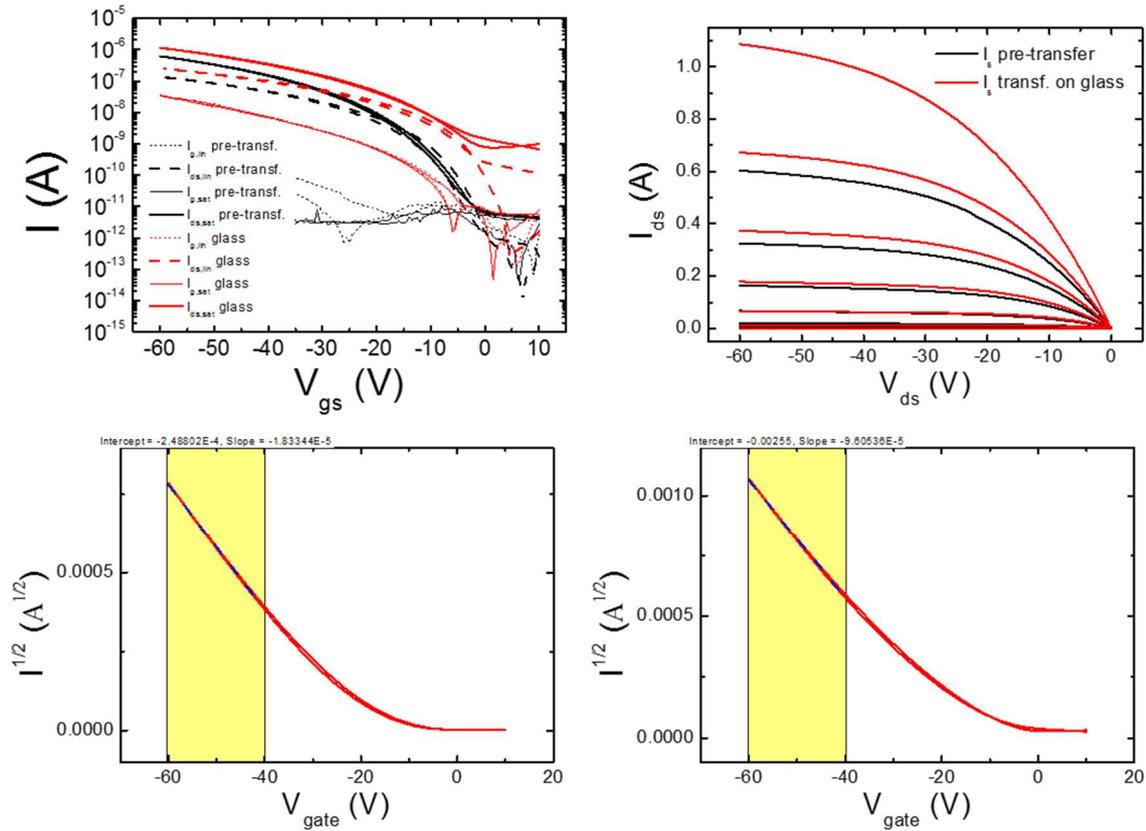

**Figure S5.** Sample OFET fabricated with P3HT:PS and transferred on glass. Transfer curves (top-left) obtained with $V_{ds,lin}$ = -5 V and $V_{ds,sat}$ = -60 V. Output curves (top-right) curves obtained with $V_{gs}$ steps of -10 V between 0 V and -60 V. Bottom images represent $\sqrt{I_{ds}}$ curves for the extraction of mobility and threshold voltage, before (left) and after (right) transfer.

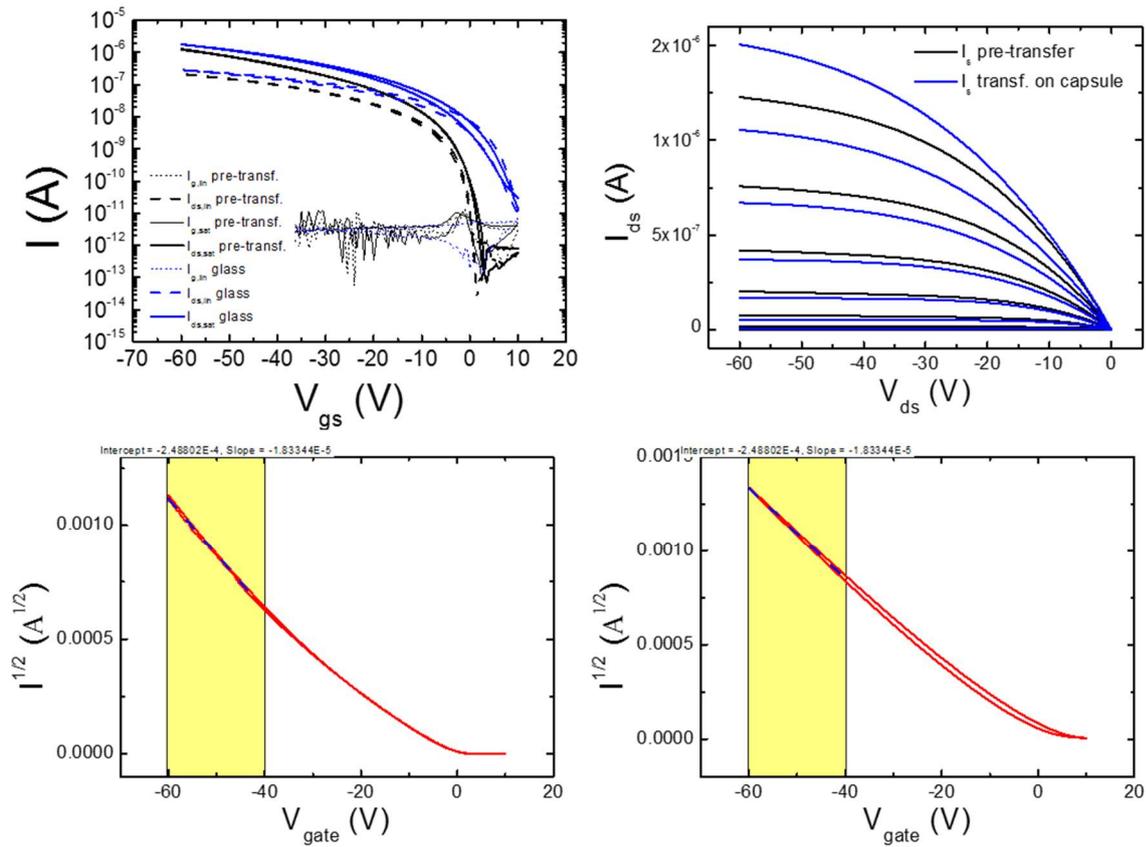

**Figure S6.** Sample OFET fabricated with P3HT:PS and transferred on capsule. Transfer curves (top-left) obtained with $V_{ds,lin}$ = -5 V and $V_{ds,sat}$ = -60 V. Output curves (top-right) curves obtained with $V_{gs}$ steps of -10 V between 0 V and -60 V. Bottom images represent $\sqrt{I_{ds}}$ curves for the extraction of mobility and threshold voltage, before (left) and after (right) transfer.

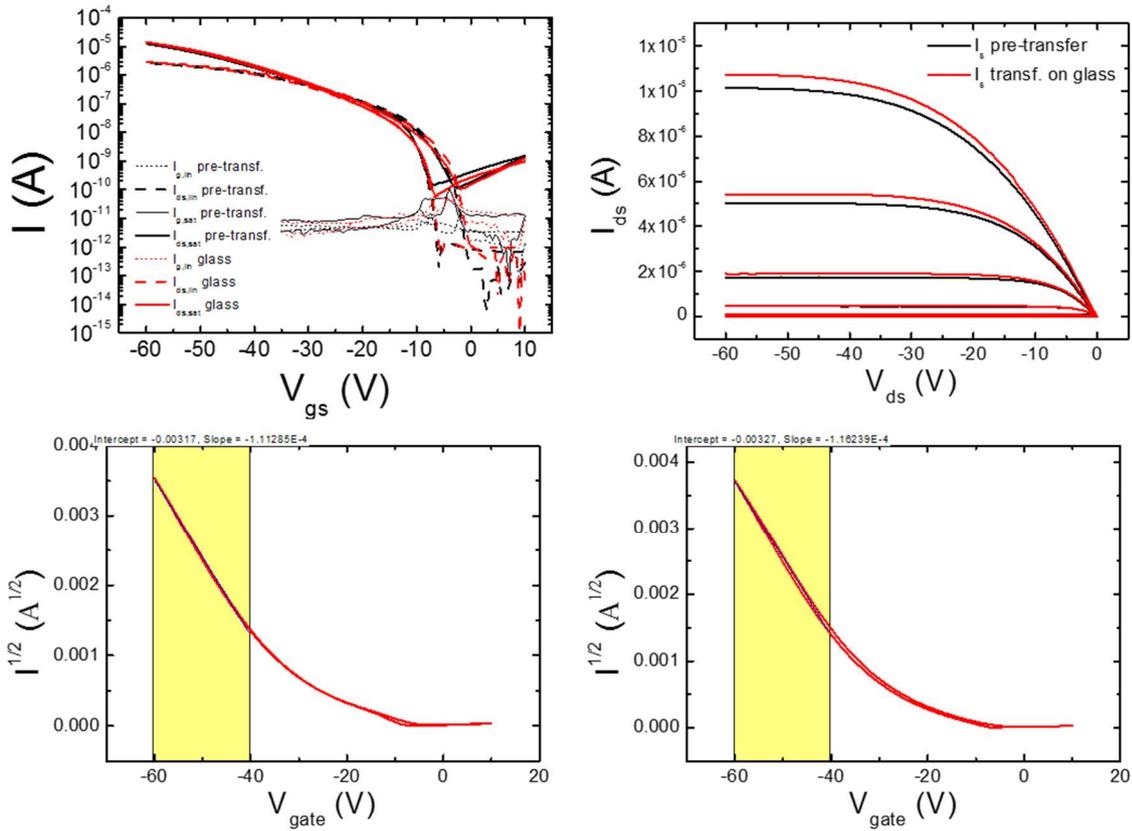

**Figure S7.** Sample OFET fabricated with 29-DPP-TVT and transferred on glass. Transfer curves (top-left) obtained with $V_{ds,lin}$ = -5 V and $V_{ds,sat}$ = -60 V. Output curves (top-right) curves obtained with $V_{gs}$ steps of -10 V between 0 V and -60 V. Bottom images represent $\sqrt{I_{ds}}$ curves for the extraction of mobility and threshold voltage, before (left) and after (right) transfer.

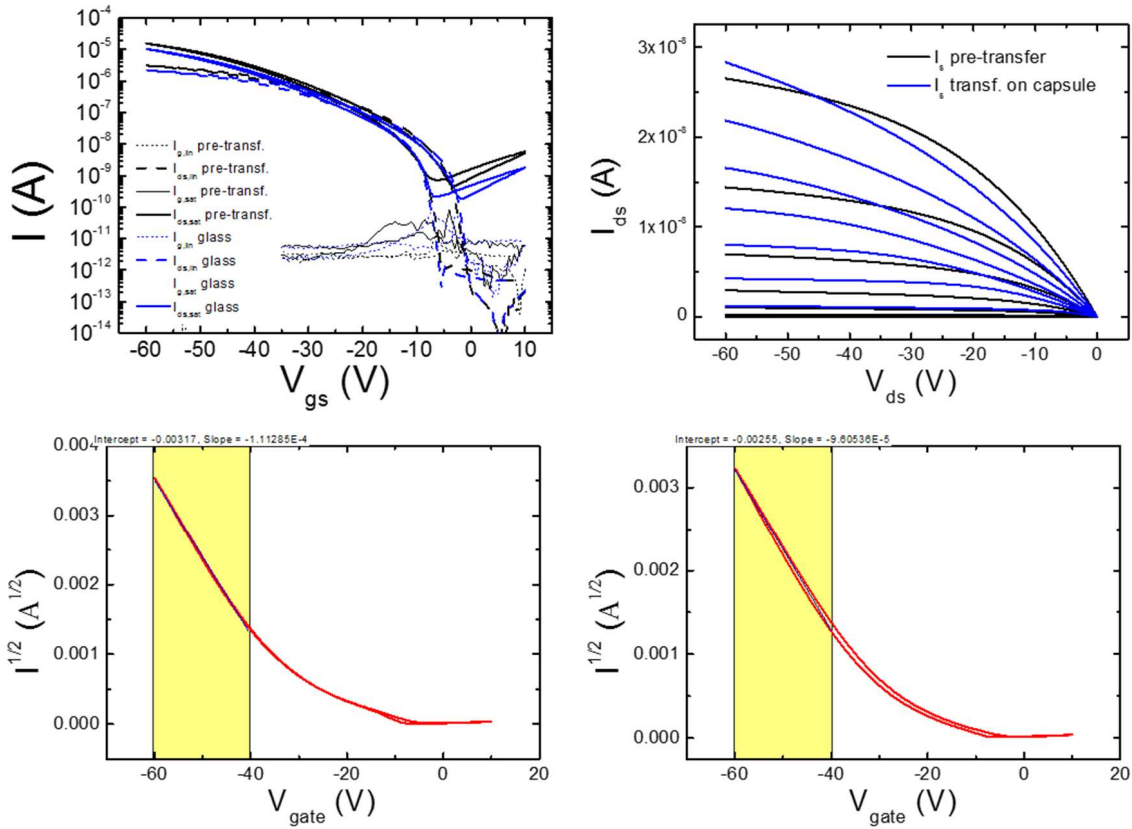

**Figure S8.** Sample OFET fabricated with 29-DPP-TVT and transferred on capsule. Transfer curves (top-left) obtained with $V_{ds,lin}$ = -5 V and $V_{ds,sat}$ = -60 V. Output curves (top-right) curves obtained with $V_{gs}$ steps of -10 V between 0 V and -60 V. Bottom images represent $\sqrt{I_{ds}}$ curves for the extraction of mobility and threshold voltage, before (left) and after (right) transfer.

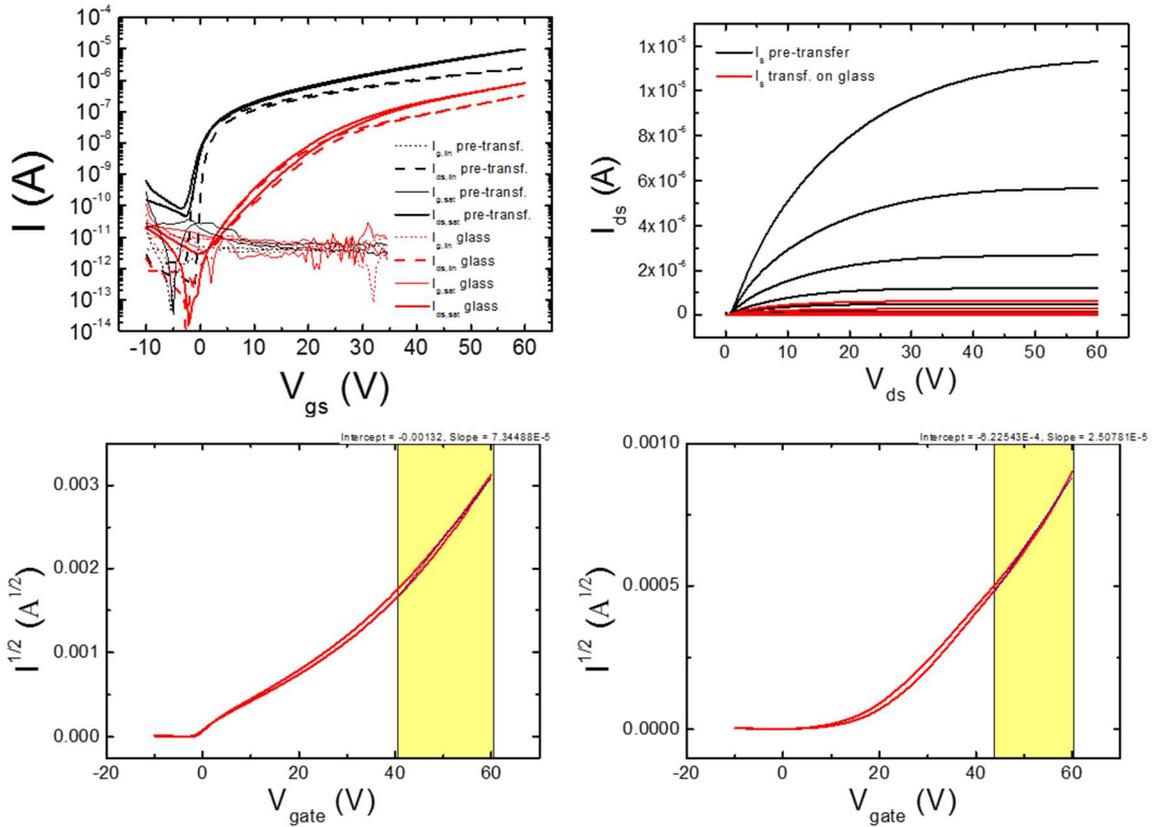

**Figure S9.** Sample OFET fabricated with P(NDI2OD-T2) and transferred on glass. Transfer curves (top-left) obtained with $V_{ds,lin}$ = 5 V and $V_{ds,sat}$ = 60 V. Output curves (top-right) curves obtained with $V_{gs}$ steps of 10 V between 0 V and 60 V. Bottom images represent $\sqrt{I_{ds}}$ curves for the extraction of mobility and threshold voltage, before (left) and after (right) transfer.

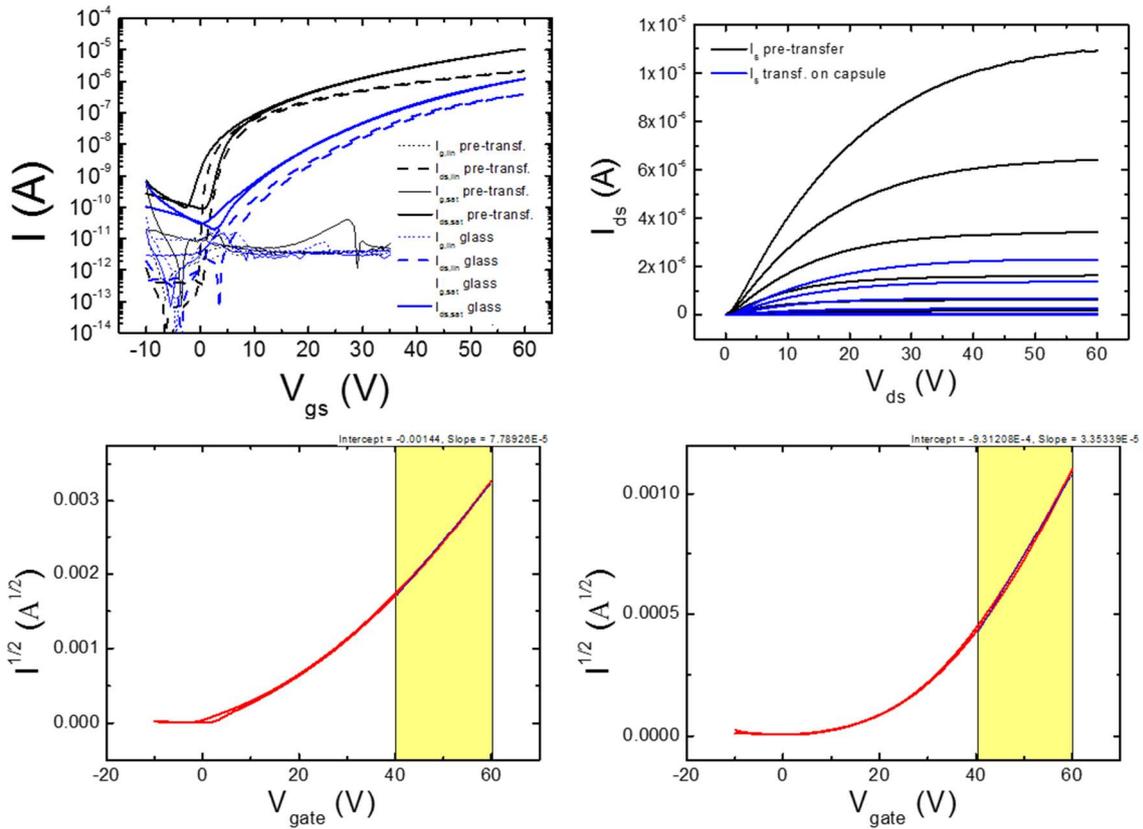

**Figure S10.** Sample OFET fabricated with P(NDI2OD-T2) and transferred on capsule. Transfer curves (top-left) obtained with $V_{ds,lin}$ = 5 V and $V_{ds,sat}$ = 60 V. Output curves (top-right) curves obtained with $V_{gs}$ steps of 10 V between 0 V and 60 V. Bottom images represent $\sqrt{I_{ds}}$ curves for the extraction of mobility and threshold voltage, before (left) and after (right) transfer.

|  | $V_{dd}$ [V] | $V_{ss}$ [V] | $V_m$ [V] | $V_{il}$ [V] | $V_{ol}$ [V] | $V_{ih}$ [V] | $V_{oh}$ [V] | $NM_l$ [V] | $NM_h$ [V] | $NM_l$ [%] | $NM_h$ [%] | Gain |

| | | | | | | | | | | | |
|---|---|---|---|---|---|---|---|---|---|---|---|
| Pre-transfer to glass | 60 | 0 | 14.5 | 7 | 3.9 | 17.5 | 56.3 | 3.1 | 38.8 | 10.3 | 129.3 | 20 |
| Pre-transfer to capsule | 60 | 0 | 17 | 11.5 | 3.7 | 20 | 56.3 | 7.8 | 36.3 | 26 | 121 | 25 |
| Transfer on glass | 60 | 0 | 38.5 | 33.5 | 4 | 43.5 | 55.7 | 29.5 | 12.2 | 93.3 | 40.7 | 15 |
| Transfer on capsule | 60 | 0 | 37 | 32 | 4 | 41 | 56.2 | 28 | 15.2 | 93.3 | 50.7 | 25 |

**Table S4.** Complementary inverter parameters, before and after the transfer. Definition of the parameters as in Sedra, A. S. & Smith, K. C. *Microelectronic circuits,* Oxford Univ. Press, Oxford, **1998**.

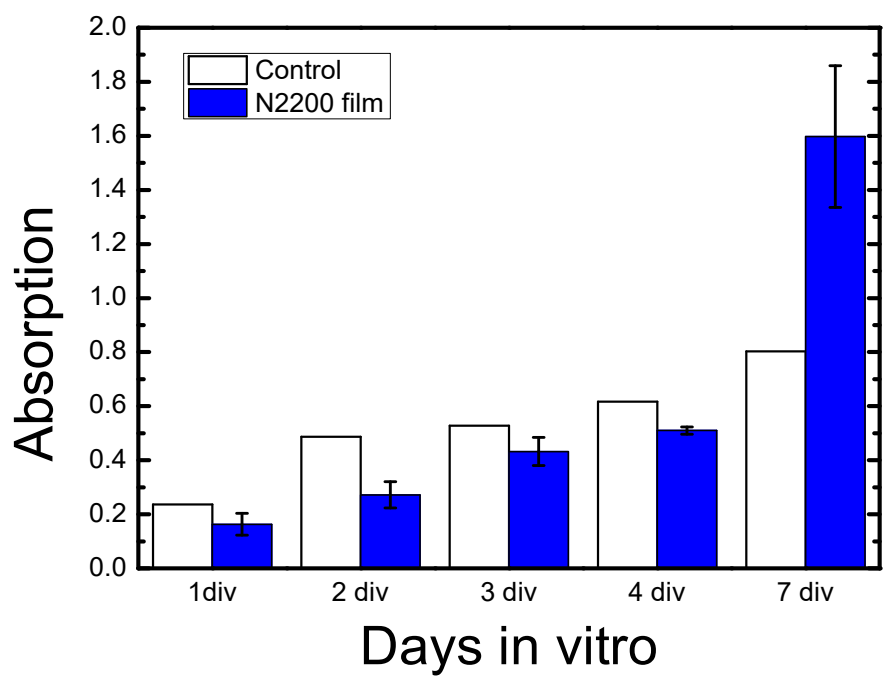

**Figure S11.** Cell proliferation assay (MTT test) performed on P(NDI2OD-T2) with human embryonic kidney cells (HEK-293). Error bars represent Standard Error estimated on three samples.

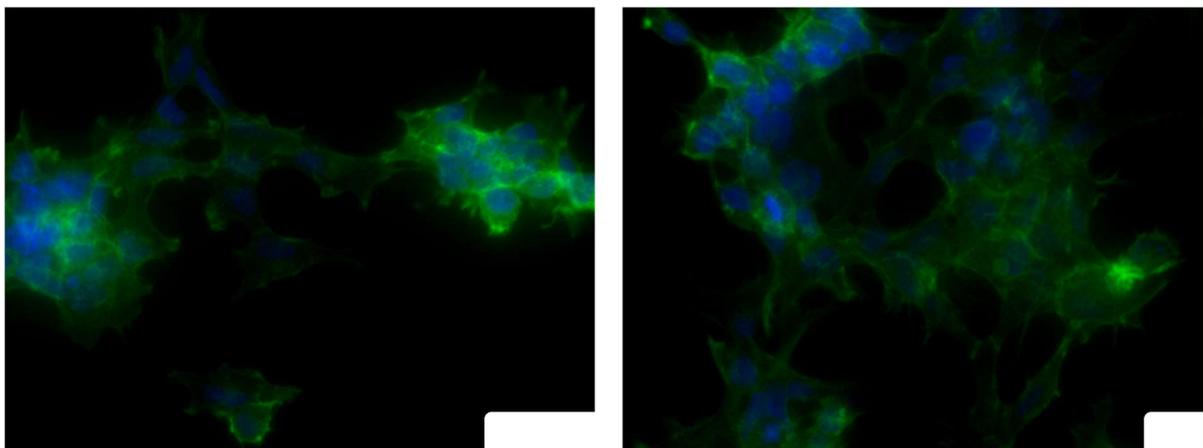

**Figure S12.** Fluorescence microscope images reported for HEK-293 cell line grown on glass control (left) and P(NDI2OD-2T) (right). F-actin proteins appearing in green and nuclei in blu. Scale bar correspond to 100 μm.

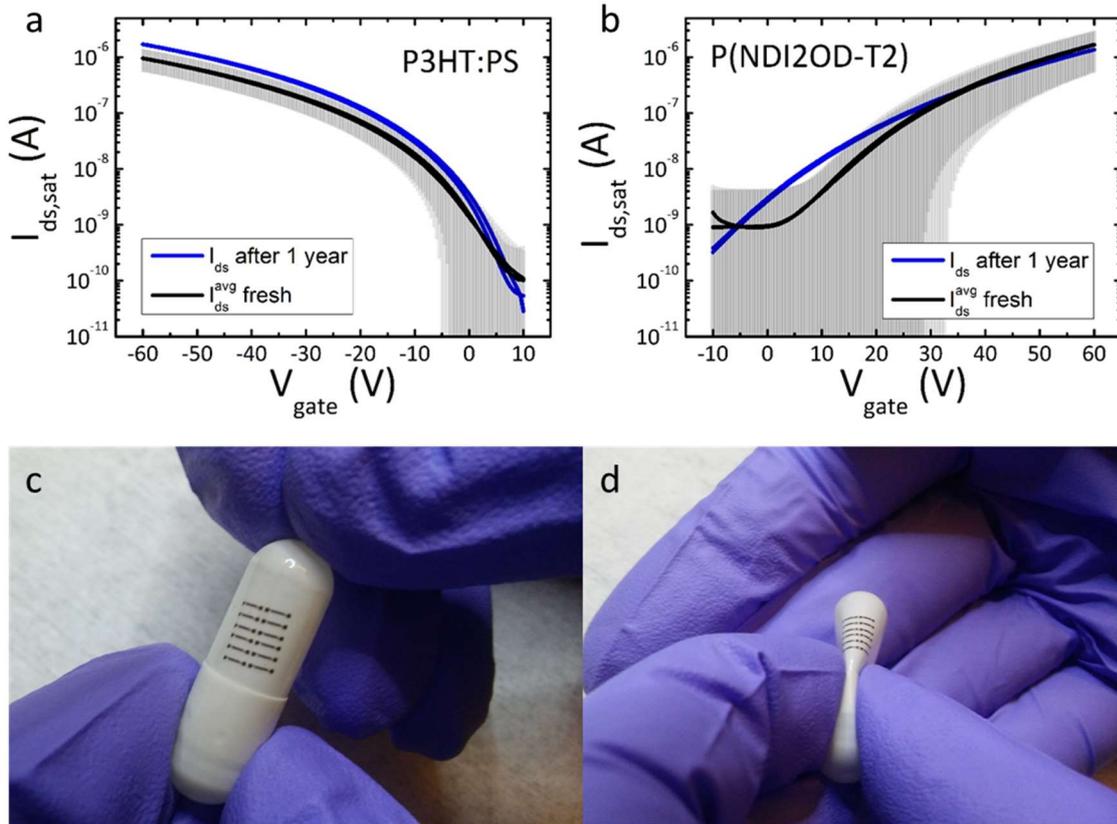

**Figure S13.** Top graphs show the transfer curves in saturation regime of one p-type (a) and one n-type transistors (b) after approximately one year of storage in a $N_2$ filled glovebox. In particular, $I_{ds}$ in saturation regime for a single 1-year-old OFET (blue) is compared with the average current (black, shaded area represents Standard Deviation) of approximately 15-20 fresh devices. Bottom images show a set of electrodes transferred on a pharmaceutical capsule approximately 1 year before the pictures were taken. After one year the tattoo has not detached from the capsule, and it does not seem to delaminate upon mechanical deformation of the capsule.